\documentclass[twocolumn]{article}

\usepackage[utf8]{inputenc}
\usepackage{authblk}
\usepackage{amsmath}
\usepackage[colorlinks=true, allcolors=cyan]{hyperref}
\usepackage{graphicx,xcolor}

\date{January 2022}

\title{\LARGE \bf Temporal quality of post-compressed pulses at large compression factors}

\author[1,*]{Esmerando Escoto}
\author[1,2]{Anne-Lise Viotti}
\author[1]{Skirmantas Alisauskas}
\author[1]{Henrik T\"unnermann}
\author[1]{Ingmar Hartl}
\author[1,3,4]{Christoph M. Heyl}

\affil[1]{Deutsches Elektronen-Synchrotron DESY, Notkestra{\ss}e 85, 22607 Hamburg, Germany}
\affil[2]{Department of Physics, Lund University, P.O Box 118, SE-221 00 Lund, Sweden}
\affil[3]{Helmholtz-Institute Jena, Fr\"obelstieg 3, 07743 Jena, Germany}
\affil[4]{GSI Helmholtzzentrum f\"ur Schwerionenforschung GmbH, Planckstra{\ss}e 1, 64291 Darmstadt, Germany}
\affil[*]{Corresponding author: esmerando.escoto@desy.de}

\begin{document}

\twocolumn[
  \begin{@twocolumnfalse}
    \maketitle
    
\begin{abstract}
Post-compression of ultra-short laser pulses via self-phase modulation is routinely employed for the generation of laser pulses with optical bandwidths reaching far beyond the laser gain limitations. While high compression factors can be routinely achieved, the compressed pulses typically suffer from temporal quality degradation. We numerically and experimentally analyze the deterioration of different measures of temporal quality with increasing compression factor and show how appropriate dispersion management and cascading of the post-compression process can be employed to limit the impact of this effect. The demonstrated saturation of pulse quality degradation at large compression factors puts novel femtosecond laser architectures based on post-compressed pico- or even nanosecond laser systems in sight.   
\end{abstract}
\vspace{5 mm}
\end{@twocolumnfalse}
  ]
\section{Introduction}
\label{sec:intro}
Femtosecond laser pulses are nowadays used for a wide range of applications \cite{mourou2019nobel,keller2003recent}. Their production typically relies on the generation inside a mode-locked oscillator, followed by amplification and optionally frequency conversion or post-compression. The latter enables the temporal compression of amplified laser pulses, surpassing the bandwidth limits of amplifier gain media. The wide spectra required to support ultra-short pulses can be generated by taking advantage of the optical Kerr effect, causing an intensity-dependent refractive index. This nonlinear effect in turn results in the modulation of the pulse's temporal phase depending on its amplitude shape and thus to spectral broadening, a phenomenon called self-phase modulation (SPM). 
While post-compression methods are widely employed for decades, the quest for ultra-short pulses at high peak and average power has pushed post-compression technology to increasing compression factors. In particular, the invention of the hollow-core capillary/fiber, followed by very recent developments employing multi-pass cells have opened the possibility to post-compress multi-100\,fs or even picoseond pulses at millijoule pulse energy to durations reaching the few-femtosecond regime \cite{nagy2021high}. Industry-grade, high-power picosecond ytterbium (Yb)-based lasers can thus be directly converted into femtosecond sources, making it possible to extend their high repetition rate, power-scalability and reliability to ultrafast strong field physics \cite{schulte2016nonlinear,balla2020postcompression,kaumanns2018multipass}.
Compressing the input pulse to less than a tenth of its original duration in a single compression stage has been demonstrated multiple times \cite{balla2020postcompression,kaumanns2018multipass, kramer2020enabling, viotti202160, kaumanns2021spectral, song2021generation, nagy2019generation, nagy2011optimal, jeong2018direct,chen2018compression, nagy2020generation}, with a recent record approaching a compression factor of 40 in a single compression stage \cite{balla2020postcompression}. 

A common drawback of post-compression is the degradation of the pulse's temporal quality induced by modulations of the spectral amplitude and higher-order phase contributions, which typically remain after compression. These effects cause pre- and post-pulses, which can easily reach more than 10\% of the peak power of the main pulse \cite{viotti2021temporal}. 
A few techniques have been developed to address the problem of temporal quality degradation in post-compression including input pulse shaping \cite{parmigiani2006ultra} and pulse cleaning methods. In particular, third-order nonlinear effects that preserve the fundamental wavelengths, such as nonlinear elliptical rotation (NER) or cross-polarized wave generation (XPW), have been demonstrated before as effective pulse cleaning tools. In fact, NER can be used in the same spectral broadening setup, to simultaneously broaden the spectrum while improving the temporal contrast \cite{khodakovskiy2019generation,seidel2021ultrafast}. 
However, NER mostly affects the contrast of the initial pulse, and not the strong pre- and post-pulses which appear after removing the SPM-induced chirp. Other pulse cleaning methods rely on the usage of frequency doubling \cite{chien1993production}, saturable absorbers \cite{fourmaux2011pedestal} or simply on spectral filtering \cite{buldt2017temporal}. 
Most of these methods come with a considerable reduction of pulse energy.
Since an increase in peak power is one of the main purposes for post-compression, such techniques are not ideal, as one limits the potential increase of peak power.

In this work, we explore the link between temporal quality degradation and high compression factors. We use a generalized model for post-compression to numerically show how different measures of temporal quality inevitably degrade with increasing SPM-induced spectral broadening factor, reaching saturation if suitable dispersion management is applied. 
We then verify our key observations experimentally using gas-based post-compression of mJ-class pulses. Finally, we outline a scheme enabling to maintain good temporal quality at high compression factors utilizing cascaded spectral broadening stages.

\section{SPM broadening and temporal quality}
\label{sec:SPMtemp}
\subsection{SPM broadening}

For common post-compression scenarios with negligible impact of linear dispersion during the spectral broadening process, the post-compression process can be split into two steps. First, an initial nonlinear step where SPM-induced spectral broadening is the main physical effect, and second, a linear compression step where the chirp induced by SPM is removed/optimized. The first step changes the spectral intensity while keeping the temporal shape almost the same, while the second step changes the temporal shape while keeping the spectral intensity constant. These two steps can be combined into one equation:
\begin{multline}
E_\textrm{out}(t) = \mathcal{F}^{-1} \bigg[\mathcal{F}\big[E_\textrm{in}(t)\exp(iB|E_\textrm{in}(t)|^2)\big] \\ \times \exp\left(i\frac{\gamma}{2} (\omega-\omega_0)^2\right) \bigg],  
\label{eq:BintGDD}
\end{multline}
where $\mathcal{F}$ denotes the Fourier transform from time $t$ to angular frequency $\omega$ and $\mathcal{F}^{-1}$ is the inverse. $E(t)$ represents the complex-valued temporal pulse envelope, written as the electric field's analytic signal representation \cite{trebino2020highly}. SPM is expressed in the first exponential term, where $B$ denotes the strength of the Kerr nonlinearity experienced by the pulse. We can normalize the value of the pulse intensity $|E(t)|^2$, such that $B$ equals the phase change at the peak of the pulse in radians, typically referred to as the B-integral \cite{nagy2021high}. Linear compression, where mostly linear chirp is adjusted, is commonly done using negatively chirped mirrors or grating compressors. The linear chirp, also referred to as group-delay dispersion (GDD) can be written as a quadratic phase in the Fourier domain, denoted as $\gamma$. The variable $\omega$ refer to the angular frequency, with $\omega_0$ being the central angular frequency.

To maintain the general applicability of the model, and the corresponding findings, we keep the model as simple as possible. Depending on the post-compression technique and parameter regime used, spatio-temporal effects and other nonlinearities have to be taken into consideration. However, SPM and chirp compensation typically still play a central role. We assume that the dispersion during the SPM-induced broadening process is minimal, which is typically the case for gas-based methods and limited spectral output bandwidths as well as for techniques involving a fiber much shorter than its dispersion length. Even solid-based techniques where the medium is very thin, or where the linear dispersion is compensated e.g. via dispersive multi-pass cell mirrors \cite{schulte2016nonlinear,weitenberg2017multi} can be modeled to a good approximation by this equation. Transmission losses are also neglected, as there is no general way to model this value. 

In Eq.~(\ref{eq:BintGDD}), the time and frequency variables can be re-scaled according to the full with at half maximum (FWHM) of the input pulse intensity, $\sigma^\textrm{in}_\textrm{FWHM}$, and can then be generalized for any wavelength and pulse width, similar to the dimensionless characteristic length scales used in soliton physics \cite{schade2021scaling}. 
This normalization approach is used throughout the paper (unless stated differently as for the experimental data), defining time as $t/\sigma^\textrm{in}_\textrm{FWHM}$, frequency as $(f-f_0)\sigma^\textrm{in}_\textrm{FWHM}$, and group-delay dispersion as $\gamma/\sigma^\textrm{in}_\textrm{FWHM}{}^{2}$. 

\begin{figure}[t!]
\centering
\includegraphics[width=\linewidth]{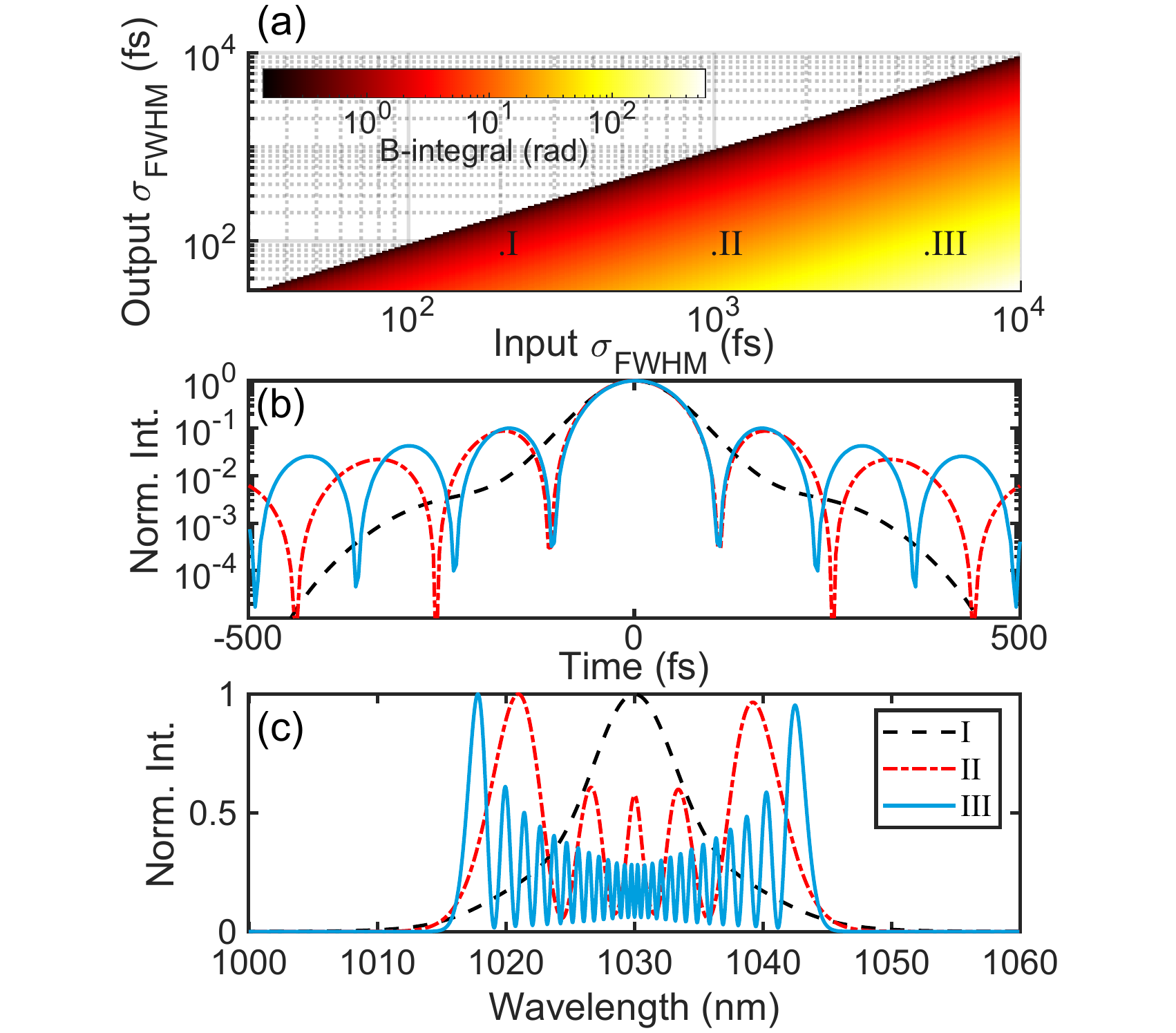}
\caption{(a) Required B-integral needed to post-compress a Gaussian pulse using SPM, based on Eq.~(\ref{eq:BintGDD}). (b) Different temporal pulse shapes of the output pulse located in the points marked by I, II, and III in (a), all three having the same FTL $\sigma_\textrm{FWHM}$ = 100\,fs. (c) Corresponding spectra of the three pulses. }
\label{fig:2d}
\end{figure}

For a Gaussian-shaped input pulse, defined by $E_\textrm{in}(t) = \exp\left(-4\ln 2\,t^2 / \sigma^\textrm{in}_\textrm{FWHM}{}^2\right)$, the temporal compression factor $C = \sigma^\textrm{in}_\textrm{FWHM}/\sigma^\textrm{out}_\textrm{FWHM}$ scales almost linearly with B-integral. Using Eq.~(\ref{eq:BintGDD}), we numerically find a suitable relation reading as 
\begin{equation}
C \approx 0.59 B + 1.
\label{eq:Capprox}
\end{equation}
This linear approximation of the compression factor as a function of the B-integral is displayed in Fig.~\ref{fig:2d}(a). It provides an approximate value valid for Fourier transform-limited (FTL) pulses as well as for pulses compressed via GDD removal only. This is because for these two cases, the functions are not exactly linear, and the two differ from the approximation in Eq.~(\ref{eq:Capprox}) almost equally except at very high compression factors, with an error below 5\% for both up to a B-integral of 40 radians. Up to 100 radians, the error is below 2\% for the FTL case and just above 10 \% with only GDD removal. 
Note that this relationship is different from the well-known relation between spectral broadening factor and B-integral, $\frac{\Delta\omega}{\Delta\omega_0} \approx 0.88B$ \cite{nagy2021high}. This means the compression factor increases slower than the broadening factor 
as the B-integral increases, an effect which can be attributed to the typical non-Gaussian amplitude shape of SPM-broadened spectra.

Figures~\ref{fig:2d}(b-c) display three post-compression scenarios yielding identical FWHM durations of the FTL pulses. With increasing compression factor, the spectral and temporal amplitude modulation period decreases and the temporal contrast deteriorates, i.e. more energy is transferred into pre- and post-pulses.  
While our simulations take into account Gaussian input pulses, it can be shown that other bell-shaped pulses behave similarly \cite{finot2018simple,boscolo2018impact}.

\subsection{Quantifying temporal quality}

There are different ways to quantify the temporal quality of an ultrashort laser pulse, depending on the property of interest. An important measure is the intensity ratio of the secondary peaks over the main peak, $I^\textrm{secondary}_\textrm{peak} / I^\textrm{primary}_\textrm{peak} $, also referred to as temporal intensity contrast. The intensity ratio matters in particular, if the pulse is used to excite matter where confinement of the interaction to one point in time can be important. As an example, laser plasma acceleration depend on laser-matter interactions at very high intensities, and pre-pulses can have a significant detrimental effect \cite{albert20212020}. An intensity ratio of 0\%  signifies that the pulse does not have any secondary peaks nor any pedestal.

Another measure describing temporal pulse quality is the \textit{energy ratio}, defined by the fraction of energy that is contained within the main peak of the pulse. This entails integrating the power over the primary peak, and dividing it by the total energy including all pre-, post-pulses and temporal pedestal. The energy ratio can be written as
\begin{equation}
    \textrm{Energy Ratio} = \frac{\int_{t_0-\sigma_\textrm{FWHM}}^{t_0+\sigma_\textrm{FWHM}}P(t)dt}{\int^{+\infty}_{-\infty}P(t)dt},
\end{equation}
where $P(t)$ is the instantaneous power of the pulse and $t_0$ is the position of the primary peak. In this equation, the boundaries of the primary peak are defined as twice the temporal FWHM width. This integration boundary is experimentally easily accessible and can be used to estimate the peak power of well-behaved SPM-broadened post-compressed pulses, as shown below. For more complicated temporal pulse structures often observed in the few-cycle regime, this boundary is less useful. An energy ratio of 100\% means that all the energy is contained within the primary peak. A perfect Gaussian pulse has an energy ratio of $\sim $98\%.

\begin{figure}[tbh]
\centering
\includegraphics[width=\linewidth]{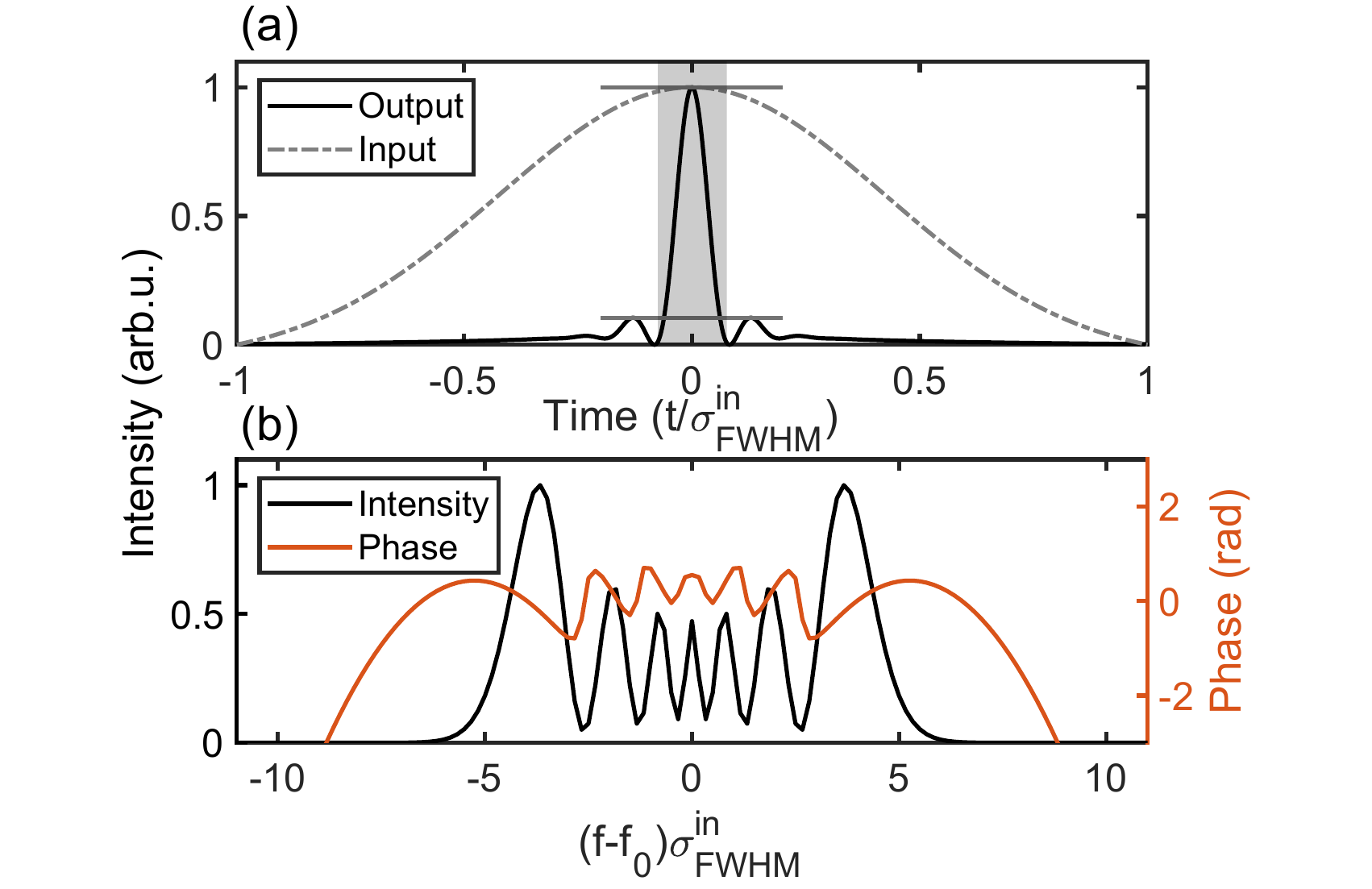}
\caption{(a) A pulse calculated from Eq.~(\ref{eq:BintGDD}), with the integration limits ($\pm \sigma^\textrm{out}_\textrm{FWHM}$) marked by the shaded area and the primary and secondary peak powers marked by horizontal line segments. (b) Corresponding spectral intensity and phase (obtained after removing the SPM-induced GDD) of the pulse.}
\label{fig:bounds}
\end{figure}

In the framework of pulse compression, another important property is the \textit{peak power boost}, which is the ratio of the peak powers after and before the compression, $P^\textrm{out}_\textrm{peak}/P^\textrm{in}_\textrm{peak}$ \cite{russbueldt2019scalable}. Note that the peak power boost can be approximated by the product of the compression factor and the energy ratio of the compressed pulse \cite{russbueldt2019scalable}. Another important factor, the energy transmission, is withheld in this discussion, as this varies greatly from case to case. 

Figure~\ref{fig:bounds} shows an example pulse that is spectrally broadened by SPM and then post-compressed according to Eq.~(\ref{eq:BintGDD}). The spectral features create secondary temporal peaks both before and after the primary peak of the pulse (see Fig.~\ref{fig:bounds}(a)). The horizontal gray line segments mark the primary and secondary peaks, highlighting the quantities used to calculate the intensity ratio. This ratio is not affected by the weak pedestal, but the energy ratio takes this into account. The shaded area has a width of twice the pulse's FWHM, serving as the integration bounds for calculating the energy ratio. The phase displayed in Fig.~\ref{fig:bounds}(b) corresponds to the higher-order phase that cannot be removed by simple GDD optimization, representing a common experimental scenario where chirped mirrors or grating-based compressors are used mainly for GDD removal.

Experimentally, pulse characterization is essential for measuring the temporal quality of the pulse. A spectrogram-based pulse shape measurement technique such as frequency-resolved optical gating (FROG) can provide the exact shape of the primary and secondary peaks \cite{trebino2020highly}. However, spectrogram-based techniques typically provide only a limited temporal dynamic range, which can lead to an overestimated peak power if a weak but possibly long pedestal cannot be resolved. For this purpose, other techniques such as third-order autocorrelation can provide an improved dynamic range  \cite{luan1993high,tavella2005high}. 

\section{Temporal quality degradation at high compression factors}

\subsection{Asymptotic degradation}

\begin{figure}[tbh!]
\centering
\includegraphics[width=\linewidth]{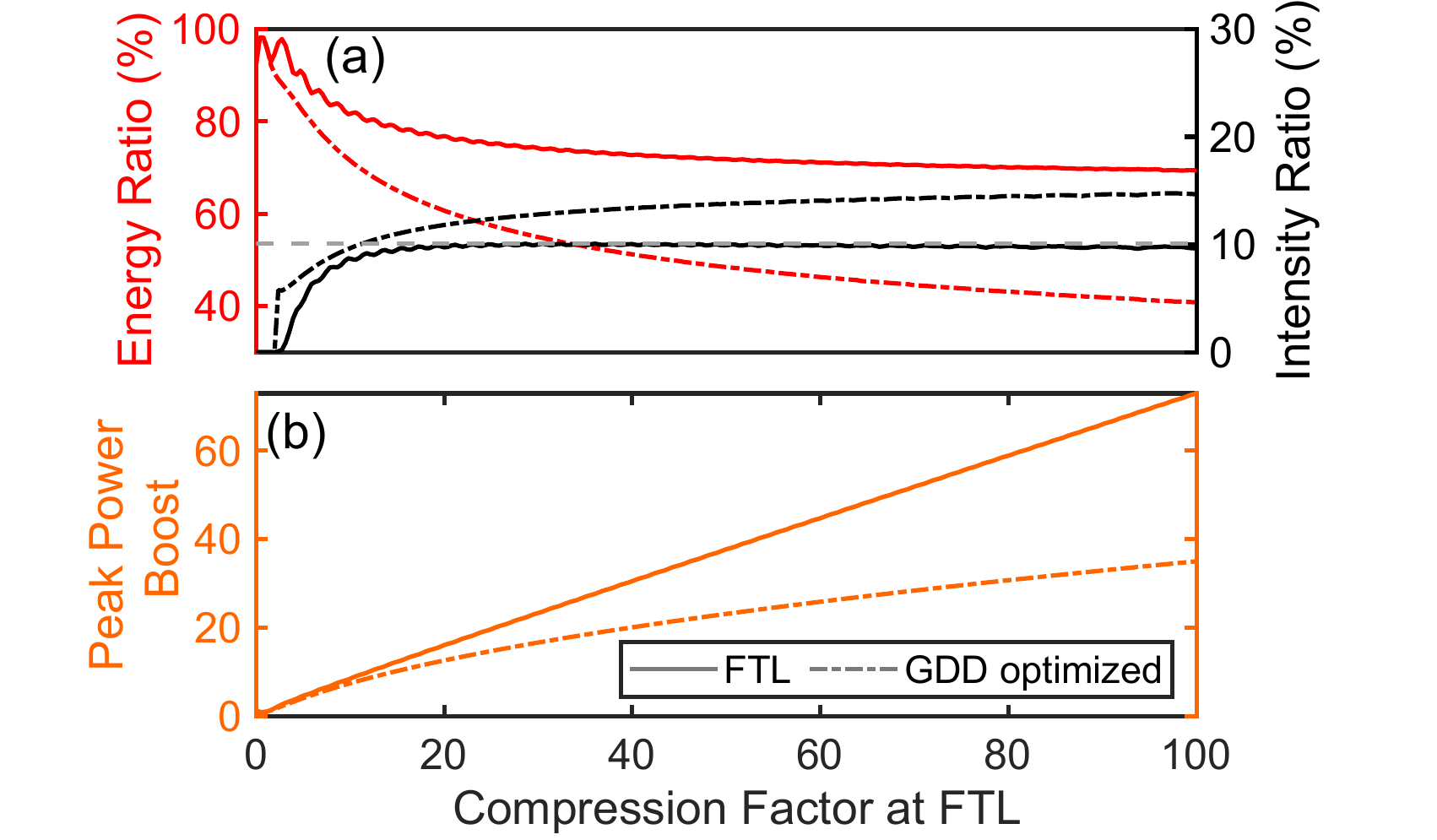}
\caption{(a) Energy ratio, intensity ratio and (b) peak power boost, plotted as functions of the compression factor at FTL. All measures are shown for two compression scenarios considering FTL (solid lines) and GDD-optimized (dashed-dotted lines) pulses.
}
\label{fig:compratio}
\end{figure}

Using the generalized treatment of pulses independent of the initial width and wavelength introduced in the previous section, we simulate post-compression of a Gaussian input pulse at different compression factors. The different measures of temporal quality are then extracted, and summarized in Fig.~\ref{fig:compratio}. 
For each measure, two cases are shown: first, the Fourier-limited case, where the SPM-induced phase has been set to zero after spectral broadening, and second, the case where only the GDD is optimized. Optimal GDD removal is determined by maximizing the peak power of the compressed pulse. 
The continuous degradation of the temporal quality with increasing compression factor can be clearly observed for the second case. However, for the FTL case, the degradation slows down and approaches an asymptotic value of 10\% for the intensity ratio (gray dashed line in Fig.~\ref{fig:compratio}(a)). The energy ratio reaches about 70\% at a compression factor of 100, showing a further slow decrease to about 65\% at a factor 1000 (not shown).

Because of the saturation effects observed for the temporal quality degradation for pulses compressed to the FTL, the peak power boost continuously increases in a linear fashion even at high compression factors. This is a clear indication that higher-order chirp becomes less tolerable at high compression factors and the need for better phase compensation techniques becomes more important. 

\subsection{Dispersion compensation}

\begin{figure}[tbh]
\centering
\includegraphics[width=\linewidth]{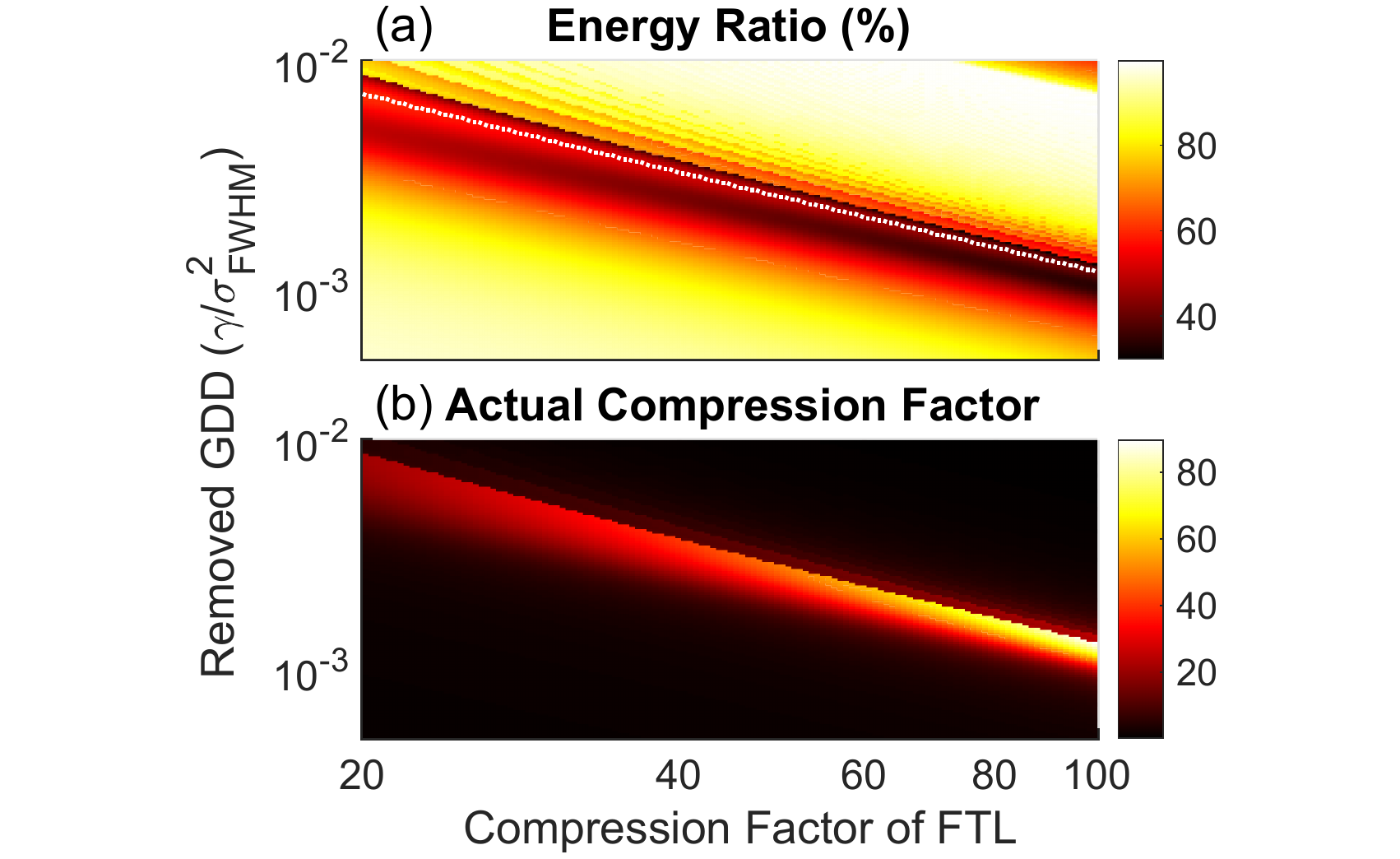}
\caption{(a) Energy ratio and (b) actual compression factor of the compressed pulses (GDD removal only). The while line in (a) marks the peak-power-optimized compression.}
\label{fig:postbroadening}
\end{figure}

We now extend the simulation results shown in Fig.~\ref{fig:compratio} to cases where the GDD is not optimally removed. The effect on the energy ratio and the actual compression factor of the compressed pulse are displayed in Fig.~\ref{fig:postbroadening}. Note that the actual compression factor refers to the ratio computed from the FWHM of the GDD-optimized pulse and not from the FWHM of the FTL pulse. 

The point of optimal compression is marked by a dashed white line in Fig.~\ref{fig:postbroadening}(a). Interestingly, the energy ratio is higher when the pulse is not optimally compressed. When the GDD is not sufficiently removed, the temporal pulse shape remains closer to the initial shape, which is a perfect Gaussian pulse. And on the other hand, when too much GDD is subtracted, multiple sub-pulses of similar height appear, resulting in a large pulse duration and consequently large integration bounds for computing the energy ratio, covering most of the pulse's energy. For very large compression ratios, the wider spectrum makes the pulse more sensitive to dispersion, and the pulse again disintegrates into multiple subpulses. These result again in a lower energy ratio, which can be seen at the top right corner of Fig.~\ref{fig:postbroadening}(a). 

A similar behavior is observed for the intensity ratio, where a better value is obtained when the pulse is not optimally compressed (not shown). Even though the energy ratio is lower when the pulse is optimally compressed, it stays above 60\% even at very high compression factors, as shown in Fig.~\ref{fig:compratio}(a).

\begin{figure}[tbh!]
\centering
\includegraphics[width=\linewidth]{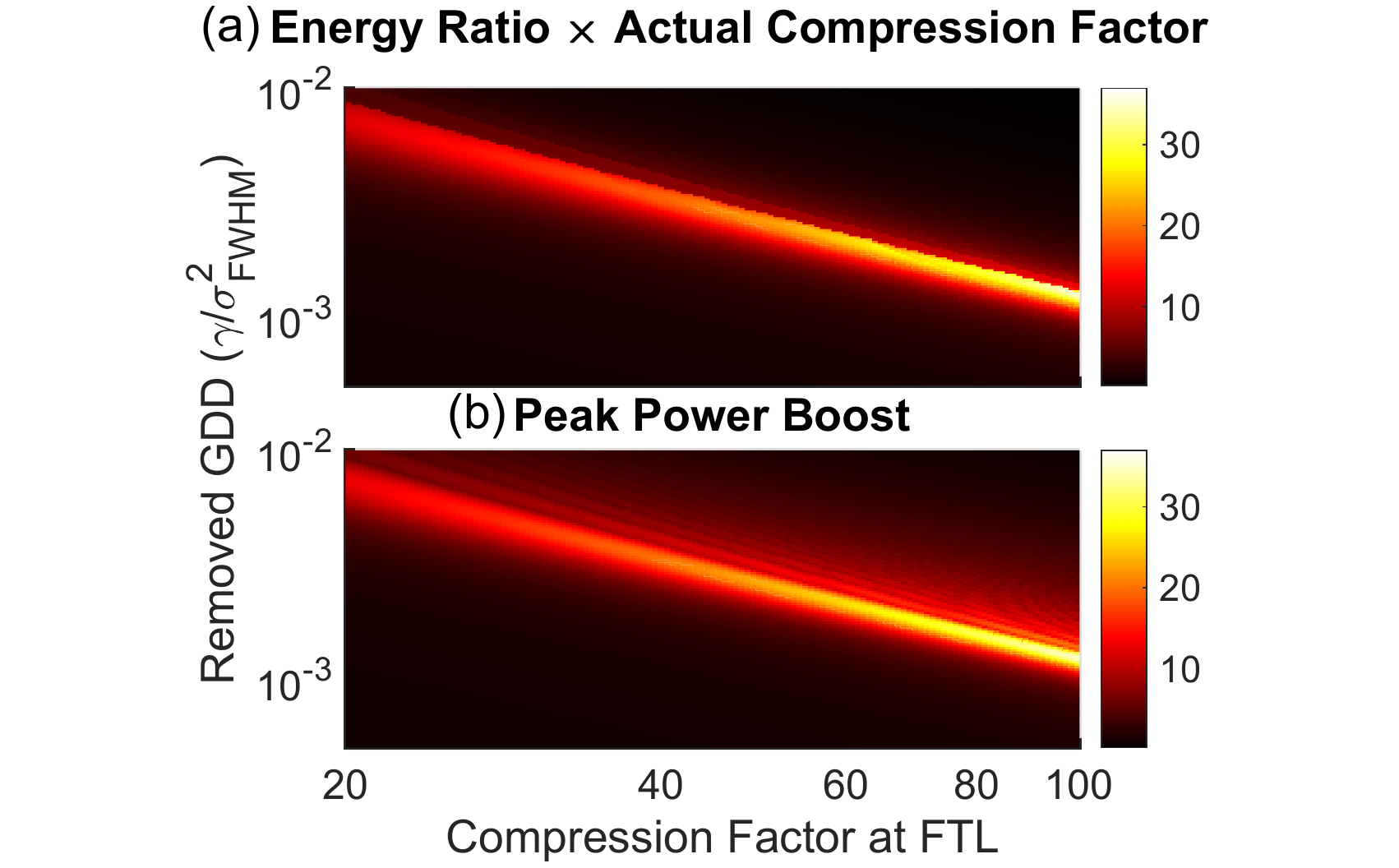}
\caption{(a) Product of the energy and actual compression factors and (b) the peak power boost. This shows the similarity of the two, showing how the product can be used as a good approximation of the peak power.}
\label{fig:compratioppb}
\end{figure}

As mentioned in Sec.~\ref{sec:SPMtemp}, the peak power can be estimated by the product of the energy ratio and the actual compression factor, as shown in Fig.~\ref{fig:compratioppb}. The figure also visualizes that even though the energy ratio gets worse at large compression factors, there is still a clear improvement of the peak power boost at an optimal GDD compensation. 

\subsection{Effect of input chirp}

\begin{figure}[tbh!]
\centering
\includegraphics[width=\linewidth]{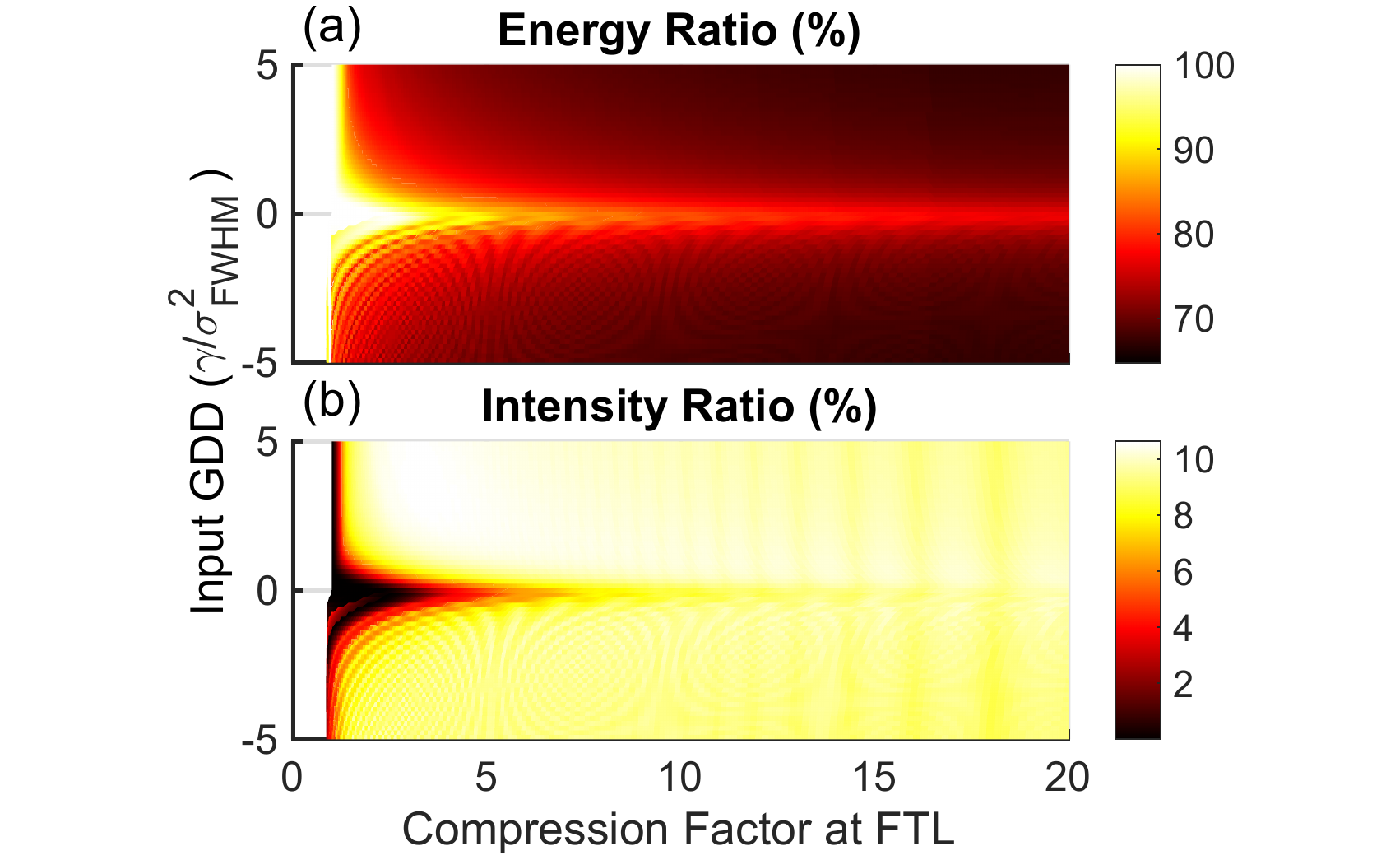}
\caption{Energy ratio (a) and the intensity ratio (b) of the compressed FTL pulse from a chirped Gaussian pulse input, plotted as functions of input pulse chirp and compression factor.}
\label{fig:gdd}
\end{figure}

Another important parameter in post-compression is the input chirp, as it can significantly alter the properties of the output pulse. One reason why chirped pulses are used in post-compression is to avoid damage due to self-focusing \cite{ganz201116fs,dombi2014conversion}. Because of the chirp, the pulse energy is distributed in a longer time span, thus lowering the peak intensity. For a Gaussian pulse, this stretching of the duration $\sigma$ can be written as \cite{diels2006ultrashort}:
\begin{equation}
    \frac{\sigma}{\sigma_\textrm{FTL}} = \sqrt{1+\left(\frac{(4\ln 2)\gamma  }{\sigma^2_\textrm{FTL}}\right)^2}. \label{eq:chirpedgaussian}
\end{equation}
Stretching the input pulse has a clear negative effect on the temporal contrast of the compressed pulse, as can be seen in Fig.~\ref{fig:gdd}. Both energy and intensity ratios degrade much faster with increasing compression factor. Note that the compression factor in this case is calculated using the FTL pulse as input pulse duration and not the duration of the stretched pulse.

Aside from the degrading temporal quality, a larger amount of nonlinearity is also needed to spectrally broaden a chirped pulse. For $\gamma/\sigma^2_\textrm{FWHM} = \pm 5$, the pulse will be almost 14 times longer according to Eq.~(\ref{eq:chirpedgaussian}). This means the peak power of the pulse will be severely diminished. For the case of negative input chirp, the spectrum initially becomes narrower at lower B-integrals as shown in \cite{daher2020spectral}, but eventually also results in spectral broadening. Note that the white part on the left hand side of the two plots do not contain data, as the compression factor is always greater than 1 when the pulse is compressing. An exemption is when the spectrum becomes narrower as mentioned, which results in a longer FTL output pulse. This is why the lower parts of the plots in Fig.~\ref{fig:gdd}, corresponding to negative input chirps, extend to compression factors below one.

\begin{figure}[tbh!]
\centering
\includegraphics[width=\linewidth]{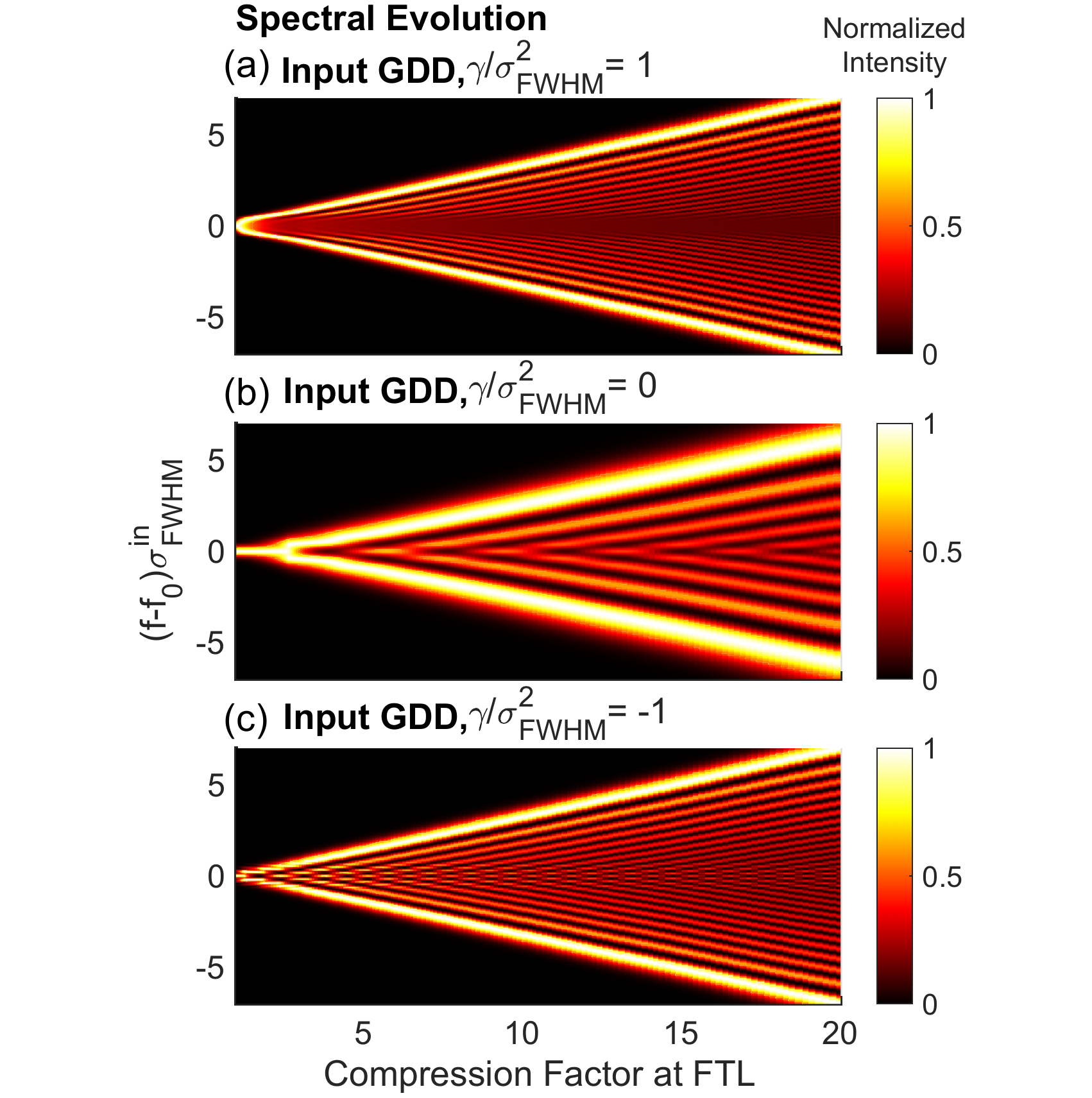}
\caption{Spectral evolution at different input dispersion values, (a) $\gamma/\sigma_\textrm{FWHM} = 1$, (b) $\gamma/\sigma_\textrm{FWHM} = 0$, and (c) $\gamma/\sigma_\textrm{FWHM} = -1$. In the first and last case, the initial pulse is stretched by dispersion to almost thrice its original duration.}
\label{fig:specev}
\end{figure}

The reduced temporal contrast for nonzero input chirp can be understood by looking at the spectral evolution of the pulse, displayed in Fig.~\ref{fig:specev}. For both positively and negatively chirped pulses, the broadened spectrum is more finely structured, which is detrimental to the temporal quality. In addition, chirped input pulses typically cause higher-order chirp under SPM broadening, which makes linear compression more difficult. These effects cause a reduced temporal quality. It is therefore not advisable to use a chirped input pulse for post-compression. Additional issues beyond temporal quality degradation caused by chirping the input pulse are discussed in Ref.~\cite{seidel2018solid}.

\subsection{Experimental demonstration of temporal contrast degradation}

For a direct demonstration of pulse quality degradation with increasing compression factor, we used a 2-m Argon-filled multi-pass cell (MPC) to compress mJ-class pulses from a Yb:YAG Innoslab burst amplifier, with an initial pulse width of 730~fs centered at 1030~nm. Third-harmonic generation FROG (THG FROG) is used to measure the pulses. Two sets of measurements were performed, one using an input pulse energy of 2 mJ and another with 3 mJ, while keeping other key parameters constant, i.e. causing an increased spectral broadening and thus a larger compression factor for the 3\,mJ case. 
The retrieved pulses have a residual GDD, which was not fully compensated in the experiment. The remaining GDD is then numerically optimized to reach optimal compression.

\begin{figure}[tbh!]
\centering
\includegraphics[width=\linewidth]{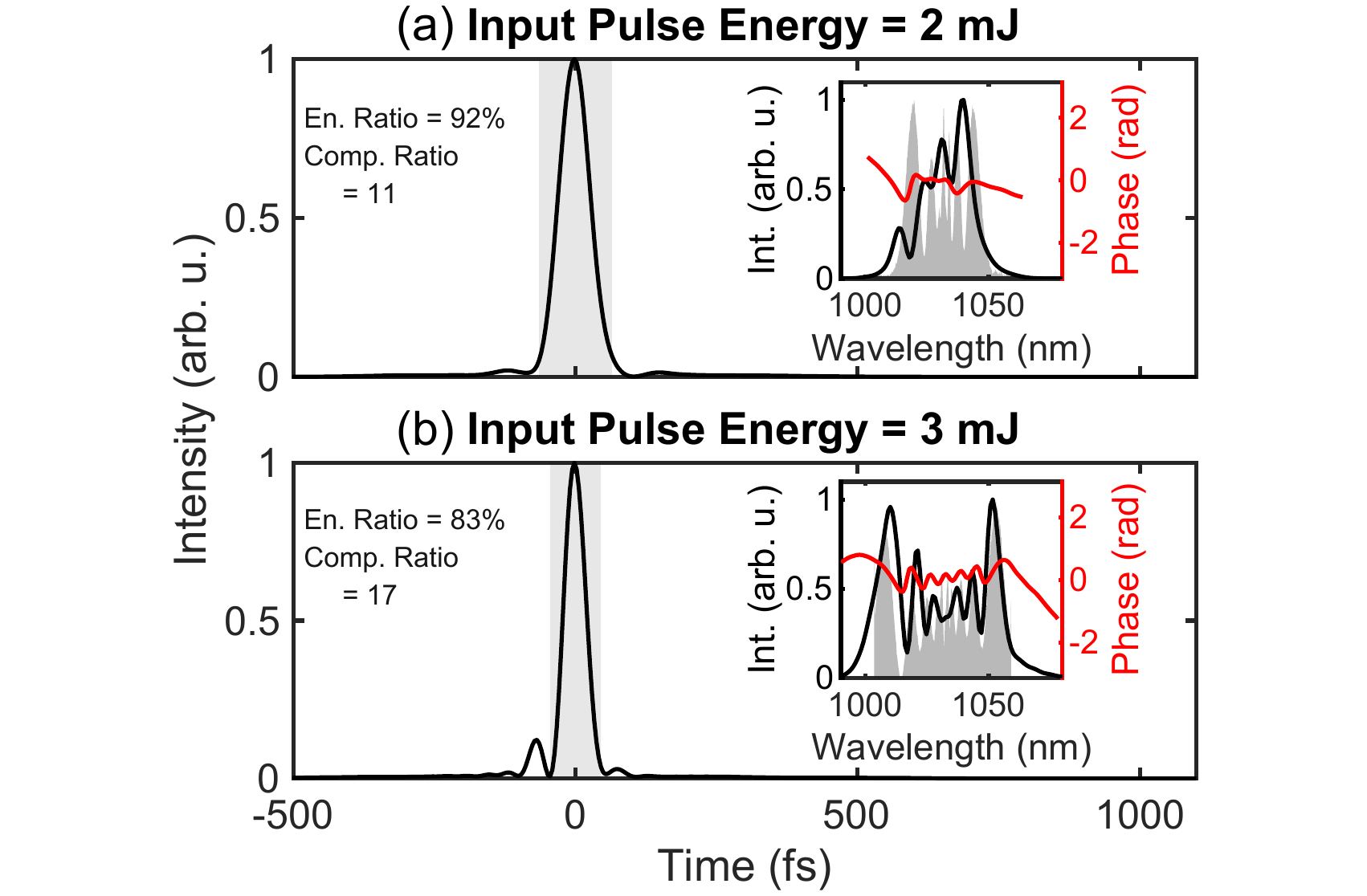}
\caption{Retrieved output pulses using an input energy of 2 mJ for (a) and an input of 3 mJ for (b). The corresponding retrieved spectral intensity and phase are plotted in the insets, with the separately recorded fundamental spectrum as shaded area.}
\label{fig:exp2pulses}
\end{figure}

The resulting reconstructed temporal pulse shapes are shown in Fig.~\ref{fig:exp2pulses}. The 2\,mJ input pulse yields an energy ratio of 92\%, which is higher than the 83\% measured for the 3\,mJ input pulse. This is expected due to the degradation of the temporal contrast at higher compression factors. Note that the dynamic range of the retrieved pulse is limited by the dynamic range of the measurement \cite{delong1994frequency}, causing most likely a slightly overestimated energy ratio for both measurements.
In addition to a reduced energy ratio, a reduced intensity ratio can be observed for the 3\,mJ case, causing stronger pre- and post-pulses. Despite this, the peak power boost is higher in the 3-mJ case, as predicted by our simulations. The peak power boost (neglecting transmission losses), is about 10 for the 2-mJ case and 14 for the 3-mJ case. 

\section{Cascaded broadening}

\subsection{Optimal cascading ratio}

Following our discussion about the degrading temporal pulse quality with increasing compression factor, we now discuss a route to mitigate this issue.  
One way to reduce the compression factor per post-compression stage is to cascade the broadening process into two stages, each with compression factors of $C_1$ and $C_2$, resulting in an effective compression factor $C_\textrm{eff}=C_1C_2$. Here, the GDD is optimized twice, first in-between the two spectral broadening stages and second, after the second stage.

\begin{figure}[tbh!]
\centering
\includegraphics[width=\linewidth]{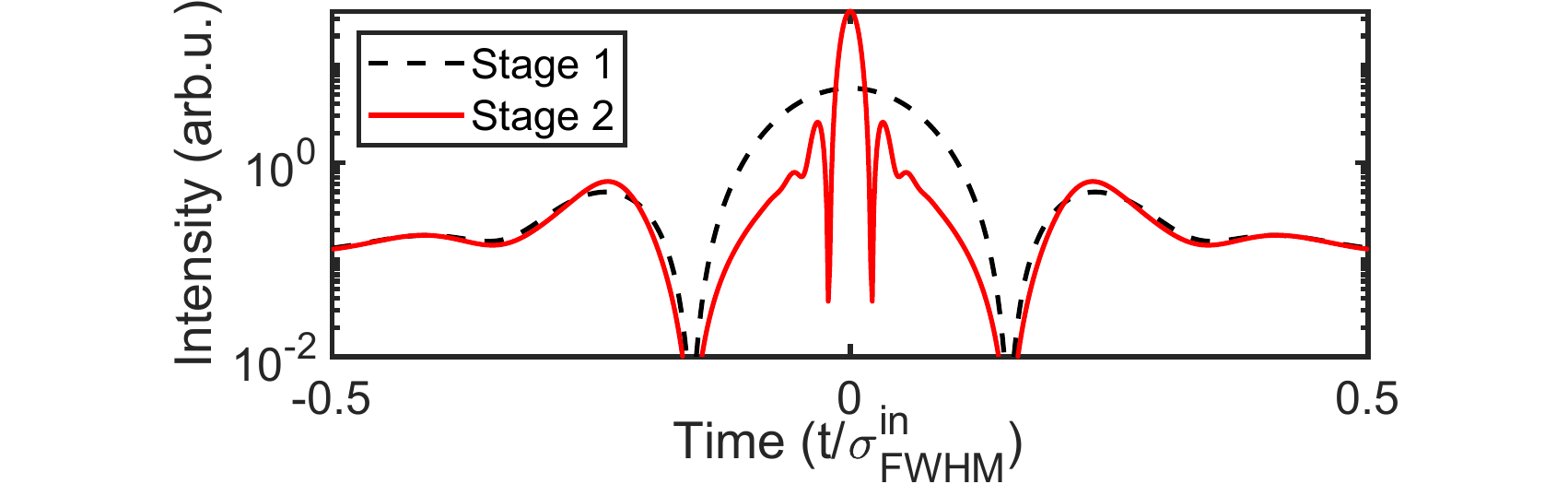}
\caption{Temporal pulse shapes resulting from a two-stage cascaded broadening process where the GDD is optimized for maximum peak power after each stage.}
\label{fig:cascade}
\end{figure}

A compressor in the middle of the spectral broadening process has the effect of halting the continuous increase of the relative peak powers of the pre- and post-pulses. Because of their reduced peak power compared to the main pulse, the pre- and post-pulses will not significantly change anymore in the next spectral broadening stage, as shown in Fig.~\ref{fig:cascade}. A very similar structure can also be clearly seen in the dual-stage compression reported in \cite{russbueldt2019scalable}. Note that up to three cascaded broadening stages have already been demonstrated \cite{fritsch2018all}.

\begin{figure}[tbh!]
\centering
\includegraphics[width=\linewidth]{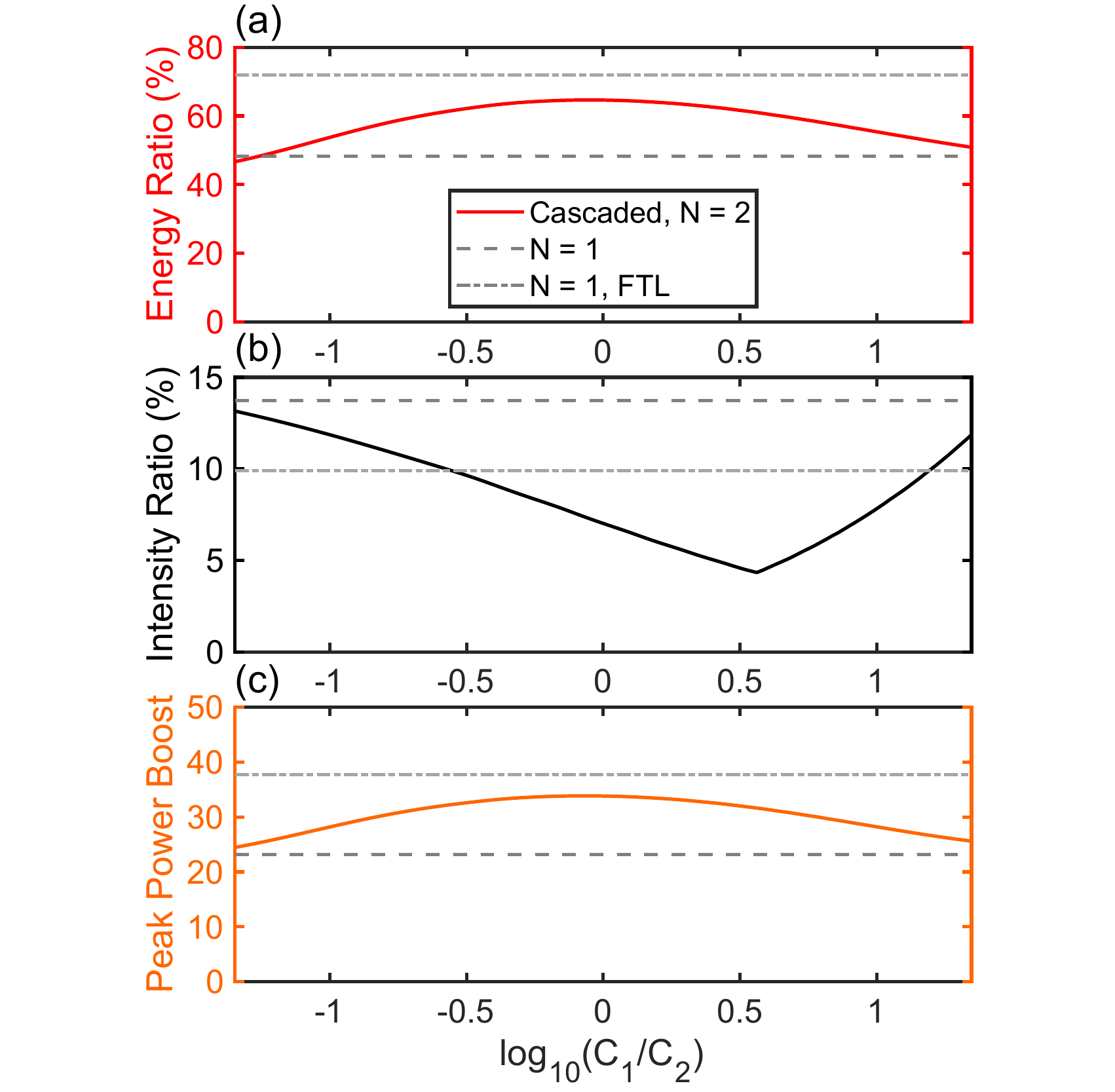}
\caption{Energy ratio (a), intensity ratio (b), and peak power boost (c) resulting from a two-stage broadening process, where $C_1$ is the compression factor of the first stage and $C_2$ for the second. In all cases, the effective compression factor $C_\textrm{eff} = C_1C_2$ is equal to 50. The dashed lines represent the case for single-stage (N = 1) broadening, where only the GDD is optimized.}
\label{fig:cascade2}
\end{figure}

In order to minimize the compression factor per stage for a given total compression factor $C_\textrm{eff}$, the optimal cascading ratio, $C_1/C_2$, should be equal to 1. This means having a compression factor for each stage of $C_i = C_\textrm{eff}^{1/N}$, where $N$ is the number of stages. To demonstrate this, we run a simulation of dual-stage compression at different cascading ratios, all resulting in the same $C_\textrm{eff}$ of 50. The results are summarized in Fig.~\ref{fig:cascade2}.
It can be seen that the optimal case for both the energy ratio and the peak power boost is the middle part of the plots, where $C_1 = C_2$. For the intensity ratio, the optimal part is not in the middle, but rather shifted towards the right side ($C_1 > C_2$). This minimum corresponds to the scenario where the pre- and post-pulses appearing in the second compression stage reach the same height as the ones generated in the first stage (see Fig.~\ref{fig:cascade}). Because the main pulses are very different in the two stages, the growth rate of these secondary peaks is also different, explaining this asymmetry.  

In general, Fig.~\ref{fig:cascade2} illustrates improved temporal quality for dual-stage post-compression compared to the single-stage case (considering compression via GDD optimization) even when the optimal cascading ratio is not met. Unless this cascading ratio is beyond 10, single-stage compression with the same compression factor $C_\textrm{eff}$ (indicated by the dashed lines) is always worse than using a dual-stage compression. 

\begin{figure}[tbh!]
\centering
\includegraphics[width=\linewidth]{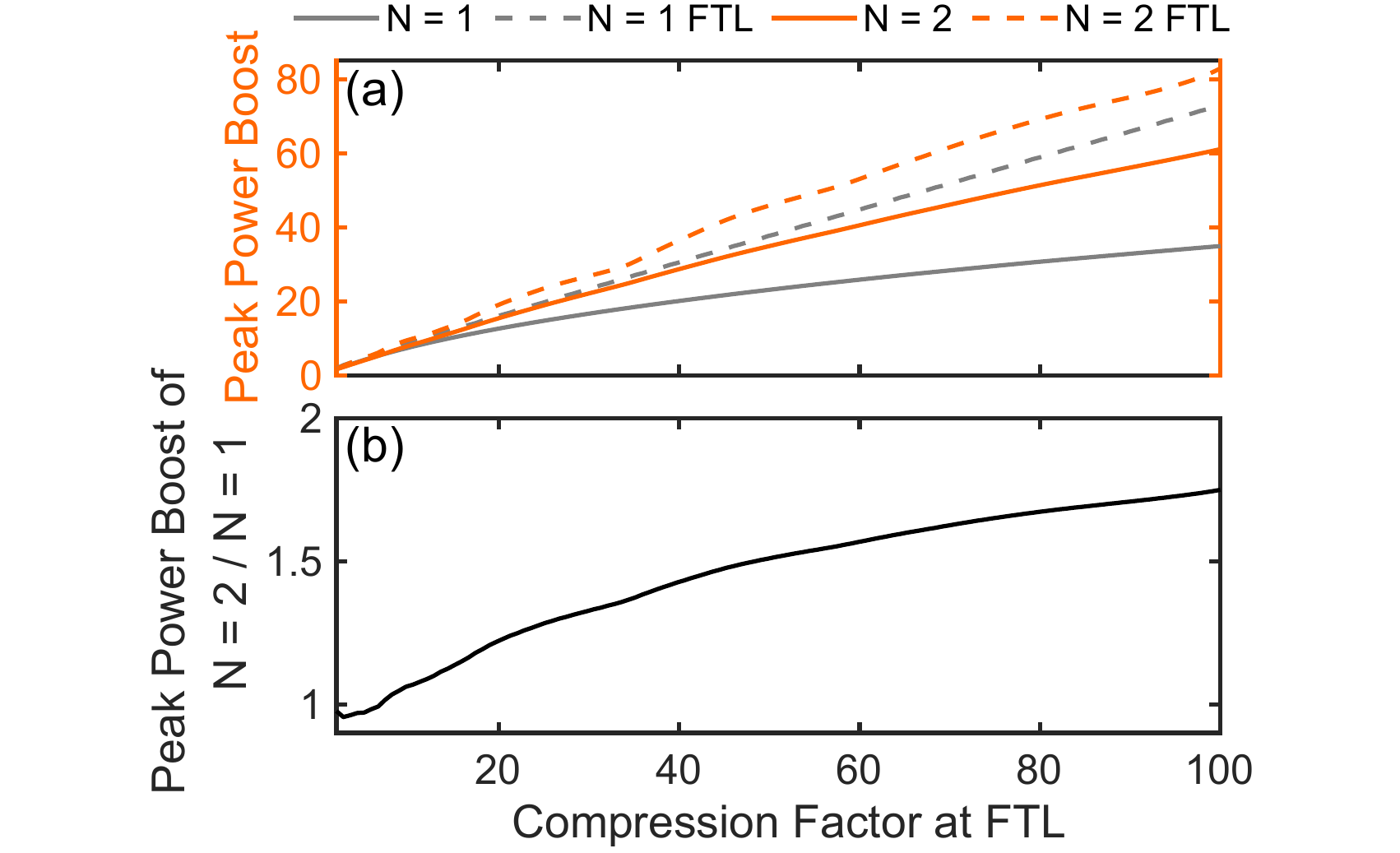}
\caption{(a) Peak power boost improvement of dual-stage cascading (N = 2) compared to single-stage (N = 1). (b) The ratio of the peak power boost for N = 2 and N = 1, where only the GDD is optimized for both.}
\label{fig:cascade3}
\end{figure}

Fig.~\ref{fig:cascade3}(a) compares the peak power boost for single and dual-stage post-compression. 
The peak power boost for the optimal dual-stage compression reaches almost the same level as the FTL case for single-stage compression. This shows that even without using advanced pulse shaping techniques to remove the higher-order chirp if large compression factors are used, dual-stage cascading enables reaching a high peak power boost. And if higher-order chirp can also be removed for the cascaded compression, an even higher peak power boost will be achieved.

An important factor which was so far neglected in our discussion about cascaded broadening is arising due to losses in particular in the compressor and possible coupling losses if fibers are employed. Naturally, the overall losses will increase with $N$. 
For MPCs, the transmission can reach above 90\% \cite{russbueldt2019scalable}, limited only by the spatial quality of the beam and the reflectance of the mirrors used. Compressor losses can also reach values below 5\%, using efficient gratings or chirped mirrors with high reflectivity. Assuming a more modest transmission of 80\% per stage, dual-stage cascaded broadening will only result in a higher peak power boost if the improvement is greater than 1/0.8 = 1.25. This is the case for compression factors above 30, as shown in Fig.~\ref{fig:cascade3}(b). If higher transmission per stage can be achieved, this limit will be moved further down to lower compression factors. 

\subsection{Multi-stage cascading}

\begin{figure}[t]
\centering
\includegraphics[width=\linewidth]{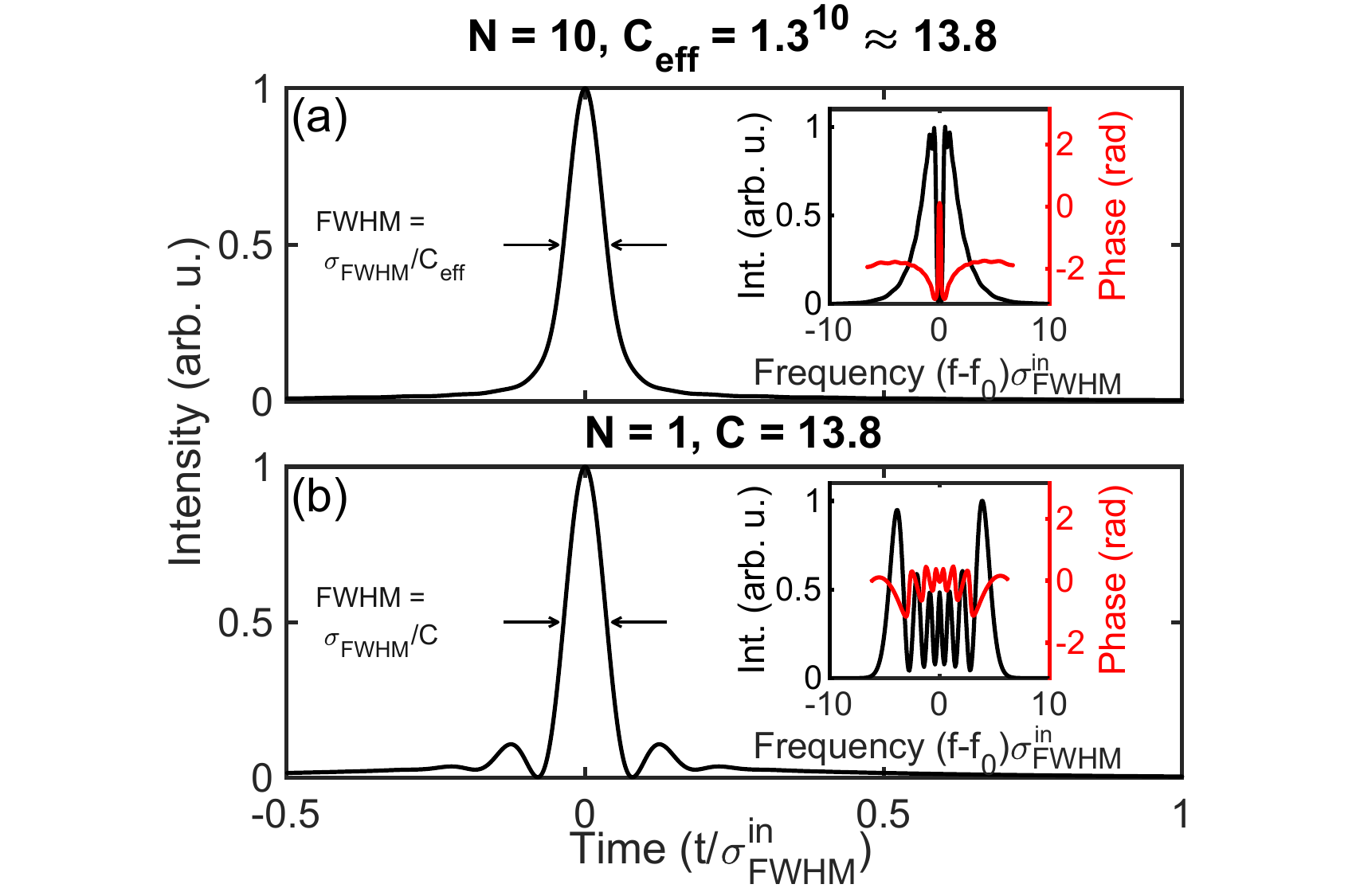}
\caption{Temporal shapes and spectra of the output pulses at (a) N = 10 and (b) N = 1, with the same input and output pulse duration for both scenarios.}
\label{fig:cascade4}
\end{figure}

As shown in Fig.~\ref{fig:cascade}, the pre- and post-pulses generated in the first stage are carried over into the subsequent stage. An interesting phenomenon occurs when the cascading is implemented before the compression factor is high enough such that these secondary pulses appear. In this case, no secondary pulse will be carried over, and the post-compressed pulse will have a very smooth profile. However, in this case, the compression factor per stage will be limited to small values, as the secondary pulses appear roughly when $C > 2$ (see Fig.~\ref{fig:compratio}). One way to still reach higher compression factors is to employ multiple cascaded stages. The result of using cascaded broadening with 10 stages is shown in Fig.~\ref{fig:cascade4}. The oscillations in the spectral intensity is less, and also the higher-order chirp is confined to a narrower region in the spectrum, overall resulting in a smoother temporal profile.

Multi-stage cascading will of course result in increased overall losses. One possible way to minimize the losses is to use an MPC, where the mirrors serve as linear compressors by employing chirped coatings, and the nonlinear medium is adjusted for each pass through the cell, e.g. by using multiple glass plates with variable thickness and/or position, such that roughly the same low compression factor is achieved for each roundtrip despite the increasing peak power. 
Reducing the B-integral per stage further and adding more stages would result in soliton-like compression with an important difference to other soliton self-comprssion scenarios: Employing an MPC results in discrete brodening and compression steps, and more importantly, the nonlinearity and dispersion per discrete step are uncoupled and can be optimized individually. Note that nonlinearity and dispersion can be adjusted in gas-filled hollow core fibers too, e.g. by using a pressure gradient, but in this case, the two properties are coupled to each other. Thus, only one is usually optimized at a time, e.g. for limiting the nonlinearity to avoid filamentation at high pulse energies \cite{bohman2010generation}.

\section{Summary and conclusion}

We showed how temporal quality degradation is connected with the compression factor of post-compressed pulses, both by using numerical simulations and with experimental data. If only the GDD of the SPM-broadened pulse is optimized, the temporal quality continuously degrades, but if higher-order chirp can also be addressed, we have shown that this degradation slows down, approaching an asymptotic limit. This conclusion is a motivation towards reaching even higher compression factors, while stressing the need for advanced phase compensation techniques to address higher-order chirp. We also showed that the temporal quality degradation in relation to higher compression factors is present regardless of the input linear chirp, but happens much faster with chirped input pulses.

We proposed the idea of using cascaded broadening specifically to limit the temporal quality degradation, especially at high compression factors. We have shown how an optimal cascading ratio can be achieved when the compression factor for each stage is minimized. Via cascading, the temporal quality of the post-compressed pulse can be improved, while at the same time reaching a higher peak power. This property highlights the importance of the discussed method as other approaches for temporal contrast improvement typically result in a lower peak power. We also showed that keeping the compression factor very low for each stage is a way towards reaching high compression factors without generating pre- and post-pulses.

In conclusion, we presented how very high post-compression factors can be achieved while minimizing temporal quality degradation by either employing higher-order dispersion management or dividing the compression into multiple stages. Our results motivate post-compression at even higher compression factors, making the technique viable for pulses in the 10-100\,ps or even the nanosecond range.


\section*{Acknowledgments \& Funding} 
We thank DESY (Hamburg, Germany) and Helmholtz-Institute Jena, members of the Helmholtz Association HGF, for support and/or the provision of experimental facilities. We acknowledge funding from the Swedish Research Council (Vetenskapsrådet 2019-06275). We also thank Marcus Seidel for fruitful discussions.

\section*{Disclosures} The authors declare no conflicts of interest.

\section*{Data availability} Data underlying the results presented in this paper are not publicly available at this time but may be obtained from the authors upon reasonable request.


\bibliography{tempcon}

\end{document}